\documentclass[12pt]{article}
\usepackage{amsmath,amsfonts,amssymb,amsthm,amstext,amscd,eucal}
\usepackage[all]{xy}

\makeatletter \@addtoreset{equation}{section}

\makeatletter\renewcommand\section{\@startsection {section}{1}{\z@}%
                                   {-3.5ex \@plus -1ex \@minus -.2ex}
                                   {2.3ex \@plus.2ex}%
                                   {\normalfont\large\bfseries}}
\renewcommand\subsection{\@startsection{subsection}{2}{\z@}%
                                     {-3.25ex\@plus -1ex \@minus -.2ex}%
                                     {1.5ex \@plus .2ex}%
                                     {\normalfont\bfseries}}

\usepackage{amsmath,amsfonts,amssymb,amsthm,amstext,bbm, amscd,eucal}
\usepackage{epsfig}
\usepackage{graphicx}
\usepackage{psfrag}
\usepackage{epstopdf}

\newcommand{\ie}{{\em i.e. }}

\newcommand{\be}{\begin{equation}}
\newcommand{\ee}{\end{equation}}
\newcommand{\bea}{\begin{eqnarray}}
\newcommand{\eea}{\end{eqnarray}}

\newcommand{\bse}{\begin{subequations}}
\newcommand{\ese}{\end{subequations}}
\newcommand{\bi}{\begin{itemize}}
\newcommand{\ei}{\end{itemize}}

\newcommand{\mpl}{M_{\rm pl}}

\def\Tr{  \mbox{Tr}   }

\begin{document}
\begin{titlepage}

\begin{flushright}\vspace{-3cm}
{\small
{\tt arXiv:} \\
\today }\end{flushright}
\vspace{-.5cm}

\begin{center}
\centerline{{\Large{\bf{Gauged M-flation After BICEP2}}}} \vspace{4mm}

{\large{{\bf A.~Ashoorioon\footnote{e-mail: a.ashoorioon@lancaster.ac.uk}$^{,a}$,
M.M. Sheikh-Jabbari\footnote{e-mail:
jabbari@theory.ipm.ac.ir}$^{,b,c}$}}}
\\


\bigskip\medskip
\begin{center}
{$^a$ \it Consortium for Fundamental Physics, Physics Department, Lancaster University, LA1 4YB, United Kingdom. }\\
\smallskip

{$^b$ \it School of Physics, Institute for Research in Fundamental
Sciences (IPM),\\ P.O.Box 19395-5531, Tehran, Iran}\\
{$^c$ \it  Department of Physics, Kyung Hee University, Seoul 130-701, Korea}\\
\end{center}
\vspace{5mm}

\end{center}
\setcounter{footnote}{0}

\date{\today}

\begin{abstract}

In view of the recent BICEP2 results [arXiv:1403.3985] which  may be attributed to the observation of B-modes polarization of the CMB with tensor-to-scalar ratio $r=0.2_{-0.05}^{+0.07}$, we revisit M-flation model. Gauged M-flation is a string theory motivated inflation model with Matrix valued scalar inflaton fields in the adjoint representation of a $U(N)$ Yang-Mills theory. In continuation of our previous works, we show that in the M-flation model induced from a supersymmetric 10d background probed by a stack of $N$ D3-branes, the ``effective inflaton'' $\phi$ has a double-well Higgs-like potential, with minima at $\phi=0,\mu$.  We focus on the $\phi>\mu$, symmetry-breaking region. We thoroughly examine  predictions of the model for $r$ in the $2\sigma$ region allowed for $n_S$ by the Planck experiment. As computed in [arXiv:0903.1481],  for $N_e=60$ and $n_S=0.96$ we find $r\simeq 0.2$, which sits in the sweet spot of BICEP2 region for $r$. We find that with increasing $\mu$ arbitrarily, $n_S$ cannot go beyond $\simeq 0.9670$. As $n_S$ varies in the $2\sigma$ range which is allowed by Planck and could be reached by the model, $r$ varies in the range $[0.1322,0.2623]$. Future cosmological experiments, like the CMBPOL, that confines $n_S$ with $\sigma(n_S)=0.0029$ can constraint the model further . Also, in this region of potential, for $n_S=0.9603$, we find that the largest isocurvature mode, which is uncorrelated with curvature perturbations, has a power spectrum with the amplitude of order $10^{-11}$ at the end of inflation. We also discuss the range of predictions of $r$ in the hilltop region, $\phi< \mu$. 

\end{abstract}

\end{titlepage}
\renewcommand{\baselinestretch}{1.1}

\section{Introduction}

Inflation, and in particular slow-roll inflation, has emerged as the leading framework to understand and explain the recent cosmological CMB observations, most notably by PLANCK mission \cite{Planck-data} and BICEP2 \cite{BICEP-data}. The most common models of inflation are those involving one or more scalars with canonical kinetic terms and a (carefully) chosen or designed potential (minimally) coupled to Einstein gravity. Despite the theoretical simplicity and addressing some theoretical issues in Big Bang early Universe cosmology, the wealth of the new observational data has made it increasingly difficult to find inflationary models which do well with data, as well as being free of theoretical issues, like naturalness, stability of inflaton potential and having a natural appearance and embedding into high energy physics models, such as beyond SM particle physics models or string theory.

Power spectrum of tensor modes, within the Einstein gravity theory, is independent of the details of the inflaton sector and sets the value of Hubble scale during inflation $H$. The recent observations of BICEP2 if  attributed to the B-mode polarization in the CMB power spectra, within the inflationary setup is generically related to the tensor modes and set $H\simeq 3-4\times 10^{-5}\mpl$, where $\mpl\equiv (8\pi G_N)^{-1/2}= 2.43 \times 10^{18}$ GeV is the reduced Planck mass.
The ratio of power spectra of tensor and scalar modes in the inflationary models with canonical kinetic terms  is $r=16\epsilon$ and the spectral tilt $n_S$ which parameterizes the scale dependence of CMB power spectrum, is $n_S-1=-6\epsilon+2\eta$, where $\epsilon$ and $\eta$ are the slow-roll parameters. Currently available data indicate that $\epsilon\sim\eta\sim 0.01$.

The simplest  inflationary models which can produce such a relatively large value $r\sim 0.2$ \cite{BICEP-data} are slow-roll large field models \cite{Mukhanov-Book}, and the most
appealing model in this family is the $\frac{1}{2}m^2 \phi^2$ chaotic model \cite{Linde-1983} with $m\simeq 6\times 10^{-6}~\mpl$. To produce these values of $r$ we generically need super-Planckian field roaming, $\Delta\phi\sim 10\mpl$, crystalized in  the Lyth bound \cite{Lyth:1996im}. The Lyth bound and $r\sim 0.2$ appear as a challenge to  supergravity and string theory motivated inflationary models in which $\phi/\mpl^2$ has generically a geometric meaning and is associated with some physical length \cite{SUGRA-inflation} which consistency of the theory requires it to be sub-Planckian.

Moreover, recent data together with number of e-folds $N_e$ required for solving cosmological flatness and horizon problems, $N_e\sim 50-60$, has other challenging implications for the inflaton potential in slow-roll large field or string theory motivated models: dimensionful parameters of the inflaton potential should be hierarchically smaller than $H$ and dimensionless parameters much much smaller than one. That is, we are in unnatural regions of the parameters space of these models and that we need to make sure the classical as well as quantum stability of the shape of the potential. The $\eta$-problem \cite{SUGRA-inflation} in string or supergravity motivated models alludes to this problem.

To ameliorate the above mentioned issues, two general ideas have been proposed: steepness of the potential and its shape may be fixed by imposing symmetries. (One cannot rely on supersymmetry, due to the fact that on an inflationary background supersymmetry is spontaneously broken at scale $H$.) The super-Planckian fields have been remedied by employing large number of fields which are abundant in supergravity or string theory setups all assisting to run forward the inflation \cite{Assisted-inflation}.\footnote{Besides the M-flation, which is the focus of this work, among string theory motivated models monodromy inflation \cite{Silverstein:2008sg} is seemingly dealing better in resolving the super-Planckian field roaming and, the steepness of the potential and $\eta$-problem by using axions as inflaton field; super-Planckian field roaming is resolved noting the periodic nature of axion field and the shape of potential is protected by the (approximate) shift symmetry of the axion potential. The original model fails to comply with the BICEP data due to small $r$ it predicts. Some variations like \cite{Hebecker:2014eua} are in a better shape with regard to compatibility with the data. This is still not clear to the authors of this paper how an axion with infinite decay constant as occurs in the above constructions are consistent with \cite{Banks:2003sx}.} Using large number of fields of almost equivalent mass, as is done in \cite{N-flation}, however, has its own drawbacks. Having $N$ number of degenerate fields will effectively reduce the cutoff scale where quantum gravity effects are expected to set in (which otherwise may be taken to be $\mpl$) is also reduced to $\mpl/\sqrt{N}$ \cite{N-reduced-Mpl,gauged-M-flation}. This will bring back the problem of super-cutoff roaming and quantum instability. Moreover for large $N$, $\sim 10^9$, light fields this will also lead to  dominance of quantum field perturbations over the classical inflaton rolling in N-flation, and hence causing a never-ending eternal inflation \cite{eternal-N-flation}.

In \cite{M-flation-1} we introduced Matrix inflation, or M-flation, which despite being a string theory motivated model conveniently bypasses the above issues.
M-flation action is derived as the low energy theory of a stack of $N$ D3-branes probing certain type IIb supergravity background with RR three-form fluxes.
This model, is hence as a $U(N)$ supersymmetric Yang-Mills theory perturbed by a mass term for three of its scalar fields which are in $N\times N$ adjoint representation of $U(N)$, as well as a cubic term for these scalars induced from the background RR-from. Although we are dealing with $5N^2$ gauge and scalar field modes, one can effectively reduce the theory to a single field sector with a double well Higgs-like inflaton potential. Of course, this theory contains a plethora/landscape of models  which effectively behave as multi-field models \cite{M-flation-2}. In this work we focus on the effective single field model. M-flation potential is protected by the supersymmetry of the bulk \cite{M-flation-1}. It does not suffer from super-cutoff field roaming and falling into the eternal inflation phase because  the spectrum of its fields around the effective single field model is not degenerate and all the modes around the classical inflationary path (which constitute $5N^2-1$ modes in gauged M-flation) have a hierarchical mass spectrum \cite{gauged-M-flation}. This hierarchical mass of isocurvature modes is the distinctive feature of M-flation compared to other multi-filed models like N-flation. This mass spectrum is such that
only a small number of the fields $N_s$, $N_s\ll N^2$, have masses below the effective cutoff scale. These are the modes contributing to the effective reduced cutoff, which is hence $\mpl/\sqrt{N_s}$ \cite{gauged-M-flation}. The reduced cutoff in (gauged) M-flation will also help to avoid the $\eta-$problem that can be generated at one loop level from the interactions of the inflaton with the graviton \cite{Ashoorioon:2011aa}. In addition, as we will see, M-flation,  does not involve hierarchically small dimensionful parameters and its dimensionless parameters are of order one.

Therefore, M-flation is a very theoretically appealing model. In this short note, we hence update our analysis of this model and show how it complies with the recent data and in particular the BICEP2. We show that unlike most of the string theory motivated inflation models, it can produce large enough tensor-to-scalar ratios as required by the data. The rest of this work is organized as follows. In section \ref{The-setup}, we review the setup of gauged M-flation and some basic analysis.  In section \ref{sec-ns-r}, we confront gauged M-flation with the current observational data. In the last section we summarize our results and make concluding remarks.

\section{Quick review of gauged M-flation}\label{The-setup}

Gauged M(atrix)-flation \cite{M-flation-1,M-flation-2,gauged-M-flation} is a sector of deformation of ${\cal N}=4$ $U(N)$ supersymmetric Yang-Mills theory coupled to 4d Einstein gravity described by the action
\be\label{action}%
 S=\int d^{4} x \sqrt{-g} \left(\frac{-M_{P}^{2}}{2} R - \frac{1}{4} \Tr(F_{\mu\nu}F^{\mu\nu})- \frac{1}{2}
\sum_{i=1}^3 \Tr  \left( D_{\mu} \Phi_{i} D^{\mu} \Phi_{i}
\right) - V(\Phi_{i}, [ \Phi_{i}, \Phi_{j}] ) \right) \, , %
\ee %
where $D_\mu$ is the gauge covariant derivative, $F_{\mu\nu}$ is the gauge field strength, and $\Phi_i, i=1,2,3$ are three $N\times N$ hermitian scalars in the adjoint representation of $U(N)$ gauge symmetry:  %
\be%
D_\mu\Phi_i= \partial_\mu \Phi_i+i g_{YM}[A_\mu,\Phi_i]\,,\qquad
F_{\mu\nu}=\partial_\mu A_\nu-\partial_\nu
A_\mu+ig_{YM}[A_\mu,A_\nu]\,,
\ee%
and the  $\Tr$ is over $N\times N$ matrices. This action which can be realized in string theory as the low energy theory of $N$ D3-branes probing certain 10b IIB plane-wave background with RR three-form field background \cite{M-flation-1}, if the potential term $V(\Phi_{i}, [ \Phi_{i}, \Phi_{j}])$ has the following form
\be\label{The-Potential}%
V= \Tr  \left( - \frac{\lambda}{4}  [ \Phi_{i},
\Phi_{j}] [ \Phi_{i}, \Phi_{j}] +\frac{i \kappa}{3} \epsilon_{jkl}
[\Phi_{k}, \Phi_{l} ] \Phi_{j} +  \frac{m^{2}}{2}  \Phi_{i}^{2}
\right),%
\ee%
with
\be\label{sugra-condition}
\lambda=8\pi g_s=2g^2_{YM}\,,\qquad \lambda m^2=4\kappa^2/9\,, %
\ee %
where $i=1,2,3$ and $g_s$ is the string coupling constant. It is notable that in terms of coordinates of stack of $N$ D3-branes $X^i$
\be%
\Phi_i\equiv \frac{{X_i}}{\sqrt{(2\pi)^3 g_s}\ l_s^2}\,,
\ee%
where $\ell_s$ is the string length.

\paragraph{Equations of motion for the scalar and gauge fields:}
\be\label{e.o.m}%
\begin{split}
& D_\mu D^\mu \Phi_i+\lambda[\Phi_j,[\Phi_i,\Phi_j]]-i\kappa \epsilon_{ijk}[\Phi_j,\Phi_k]-m^2\Phi_i=0\,,\cr
& D_\mu F^{\mu\nu}-ig_{YM}[\Phi_i,D^\nu\Phi_i]=0\,.
\end{split}
\ee
\paragraph{Reduction to effective single-field sector.}
As noted in \cite{gauged-M-flation, M-flation-1,M-flation-2}, one can consistently
restrict the classical dynamics to the SU(2) sector,
where one effectively deals with a single scalar field $\hat \phi$ and the matrices $\Phi_{i}$ are
\be \label{phiJ}%
\Phi_{i} =
\hat \phi(t) J_{i}\ , \quad \quad i=1,2,3,%
\ee%
where $J_{i}$ are the $N$ dimensional irreducible
representation of the $SU(2)$ algebra%
\be\label{J}%
 [ J_{i}, J_{j} ]=  i \, \epsilon_{ijk} J_{k} \ , \qquad \Tr (J_{i} \,
J_{j})= \frac{N}{12} ( N^{2}-1 ) \, \delta_{ij} \, .%
\ee%
Hermiticity of $\Phi_{i}$'s and $J_{i}$'s guarantees that
$\hat \phi$ is a real scalar field. One can consistently turn off the gauge fields, \ie $A_\mu=0$, in the background as the $[\Phi_i, D_\nu\Phi_i]$ term in the equation of motion of the gauge field for the ansatz \eqref{phiJ} is proportional to $[J_i,[J_i,A_\nu]]$. Therefore,
the classical inflationary trajectory takes place in the SU(2) sector of the scalar fields $\Phi_i$'s.

Plugging the ansatz into the action (\ref{action}) and adding the four-dimensional Einstein gravity, one obtains
\be S= \int
d^{4} x \sqrt {-g} \left[- \frac{M_{P}^{2}}{2} R+  \left( -
\frac{1}{2}  \partial_{\mu}  \phi  \partial^{\mu}  \phi -V_0(\phi)\right)\right] \, ,
\ee
where
\be\label{phi-scaling-V0}%
\begin{split}
\hat \phi &= \left(  \Tr
J^{2}   \right)^{-1/2} \phi = \left[ \frac{N}{4}(N^{2}-1)
\right]^{-1/2} \, \phi \, , \cr
V_0(\phi) &= \frac{\lambda_{eff}}{4} \phi^{4} -
\frac{2\kappa_{eff}}{3} \phi^{3} + \frac{m^{2}}{2} \phi^{2} \, , %
\end{split}
\ee%
with%
\be \label{lameff}%
\lambda_{eff} = \frac{2 \lambda}{\Tr J^{2}} = \frac{8 \lambda}{ N
(N^{2}-1)}  \ , \quad \kappa_{eff} = \frac{ \kappa}{\sqrt{\Tr J^{2}}} = \frac{2
\kappa}{\sqrt{N(N^{2}-1)}}\,. %
\ee%
As we will see, assuming that the parameters, $\lambda,\ \kappa$ and $m^2$, are related as in \eqref{sugra-condition}
the potential takes a ``symmetry breaking'' Higgs potential. With super-Planckian v.e.v's for the effective inflaton $\phi$, the potential is capable of realizing inflation in different regions. We note that this may happen while the ``physical field $\Phi$ and/or stringy lengths $X^i$ are sub-Planckian or sub-stringy.

\paragraph{Spectrum of gauged M-flation ``spectator'' modes.}

In the gauged M-flation, starting with three $N^2$ scalar fields
$\Phi_i$ and four $N^2$ gauge fields we have $7N^2$ degrees of freedom. Picking up the SU(2)  sector
\eqref{phiJ}, we have turned on only one configuration of these fields. The gauge fields satisfy the
the gauge symmetry of the action and equations of motion. Therefore we expect only $2N^2$ of the
them remain as physical and, aside from the mode along the SU(2) sector, in total we have $5N^2-1$ modes, which even though are classically frozen, could be excited
quantum mechanically. One can show that the backreaction of these isocurvature modes on the inflationary background  during slow-roll  period, is very small \cite{M-flation-1,M-flation-2,gauged-M-flation}. Thus they are truly the ``\textit{spectator}'' modes. Below we will present the spectrum of spectator modes.
\begin{itemize}
\item\textbf{Scalar modes:} The spectrum of scalar fluctuations could be obtained perturbing $\Phi_i$'s around the $SU(2)$ sector
\be\label{Phi-expand}%
\Phi_i=\hat \phi J_i+\Psi_i\,, %
\ee%
expanding the action to second order in $\Psi$'s and diagonalizing the second order action. One then find two class of modes \cite{DSV}:
\begin{itemize}
\item $\alpha_j$-modes: $\omega=-(j+2)$ and $0\leq j\leq N-2$. Degeneracy of each $\alpha_j$ mode is $2j+1$. Therefore there are $(N-1)^2$ of these modes. The mass of these modes are
\be%
M^2_{\alpha_j}= \frac12\lambda_{eff}\phi^2 (j+2)(j+3)-2\kappa_{eff}\phi (j+2)+m^2\,.
\ee%
 The
$\alpha_{j=0}$ mode is the quantum fluctuations of the direction along the SU(2) sector which are nothing other than the adiabatic mode. Thus we have $(N-1)^2-1$  $\alpha-$mode.

\item $\beta_j$-modes: $\omega=j-1$ and $1\leq j\leq N$. Degeneracy of each $\beta_j$-mode is $2j+1$. Therefore there are $(N+1)^2-1$ of $\beta$-modes. Mass of $\beta_j$ mode is%
\be
M^2_{\beta_j}=\frac12 \lambda_{eff}\phi^2 (j-1)(j-2)+2\kappa_{eff}\phi (j-1)+m^2\,.
\ee%

\end{itemize}

\item\textbf{Gauge modes:}
Likewise for the gauge fields, turning on the $A_{\mu}$'s and expanding the action \eqref{action} to second order in them assuming that the scalar fields are in the SU(2) sector, {\it i.e.} $\Phi_i=\hat\phi J_i$, one can obtain the mass spectrum of gauge modes solving the related eigenvalue problem \cite{gauged-M-flation}. The masses turn out to be
 \be%
M^2_{A,j}=\frac{\lambda_{eff}}{4}\phi^2 j(j+1)\,.
\ee%
where $j=0,\cdots, N-1$ and degeneracy of each mode is $2j+1$. $j=0$ mode is massless and corresponds to the $U(1)$ sector in the $U(N)$ matrices. Degeneracy of each $j$ gauge mode is  $3(2j+1)$ for $j\geq 1$ modes and is two for $j=0$ mode. Thus we have $3N^2-1$ vector field modes.

We note that there are $N^2-1$ ``zero-modes'' in the scalar sector, which are eaten up through the usual spontaneous breaking occurring due to expansion around the $SU(2)$ vacuum \cite{DSV,gauged-M-flation}.
\end{itemize}

In summary we have $(N-1)^2-1$  $\alpha$-modes, $N^2+2N$ $\beta$-modes
and $3N^2-1$ vector field modes, which altogether yields $5N^2-1$ modes, exactly the number of spectator (isocurvature) modes that we started with.

\paragraph{The UV safety of M-flation.} The spectrum of spectator modes of M-flation is hierarchical and the masses vary from $\sim \sqrt{\lambda_{eff}}\phi$ to
$\sim N\sqrt{\lambda_{eff}}\phi$. For our inflationary trajectories (\emph{cf.} next section) this is $10^{-6}\mpl$ to $10^{-1}\mpl$ and most of these $5N^2-1$ spectator modes have masses around the upper bound. Therefore, the effective UV cutoff in M-flation model is still very close to $5\times 10^{-3}\mpl$ \cite{Ashoorioon:2011aa,gauged-M-flation}. See section \ref{discussion-sec} for more discussions on this point.

\section{Predictions of M-flation in the $(n_S-r)$ plane}\label{sec-ns-r}

\begin{figure}[ht]
\centerline{\includegraphics[angle=0,scale=0.7]{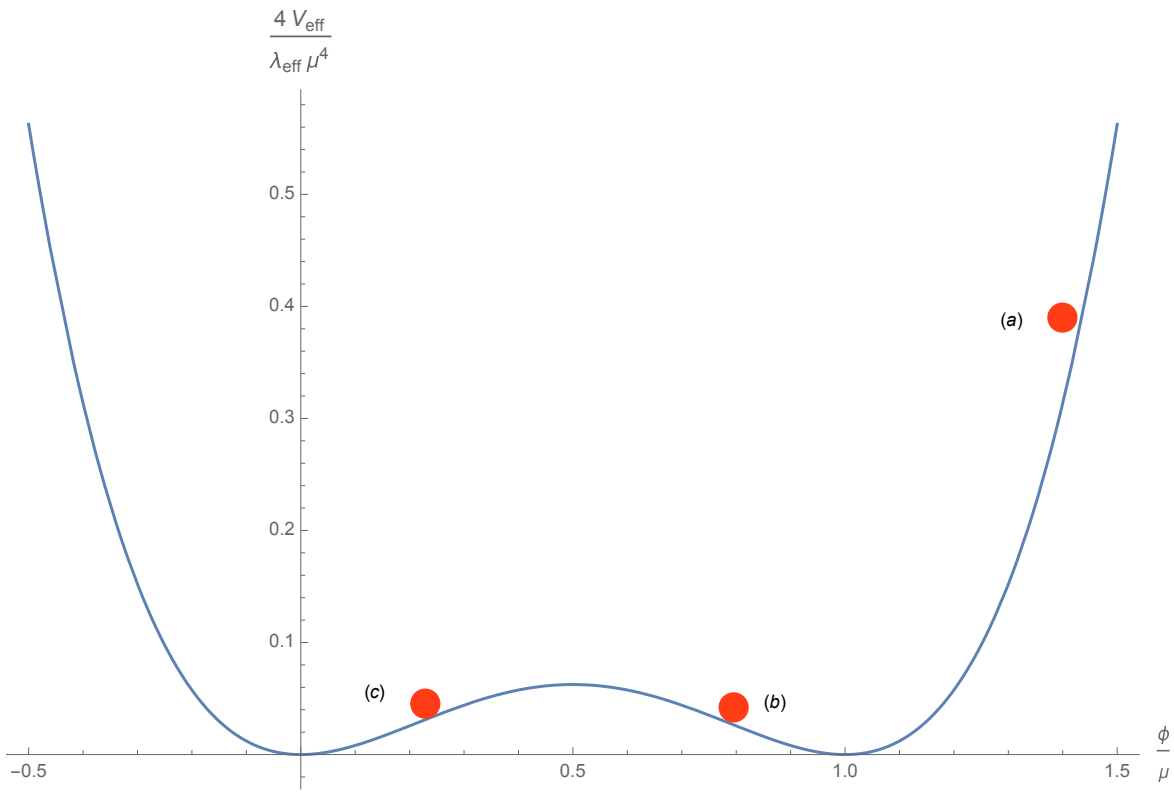} }
\caption{The double-well potential which is derived in the context of M-flation. Confining to the $\phi\geq 0$ region, there are three regions one can use to inflate upon. Due to the symmetry of the potential the predictions of the model in region (b) and (c) are the same in the $(n_S,r)$ plane. This degeneracy breaks down at the level of isocurvature perturbations. }\label{potential-shape}
\end{figure}

As noted in \cite{M-flation-1, M-flation-2}, the inflationary potential for the supersymmetric choice \eqref{susy-cond}, which reduces to the following condition in terms of effective couplings
\be\label{susy-cond}
\kappa_{\rm eff}^2=\frac{9}{8}\lambda_{\rm eff} m^2,
\ee
the potential takes a double-well form
\be\label{pot-susy}
V(\phi)=\frac{\lambda_{\rm eff}}{4}\phi^2 (\phi-\mu)^2
\ee
where
\be\label{mu}
\mu=\sqrt{\frac{2}{\lambda_{\rm eff}}}m.
\ee
The form of the potential is a Higgs-like double-well potential, Fig.\ref{potential-shape}, where, as we will notice shortly, $\mu$ has to take super-Planckian values to be able realize inflation. Such super-Planckian values for the effective inflaton could be easily accommodated within the setup of M-flation, as the physical excursion are normalized by $\sqrt{N(N^2-1)/4}$ from the roaming of the effective inflaton \eqref{phi-scaling-V0}.
In the stringy picture, under the influence of the RR six-form flux, two of the dimensions perpendicular to the D3-branes are blown up to a fuzzy sphere, $S^2$, and we have a polarized fuzzy D5 brane. The SU(2) direction will play the role of the radius of the fuzzy sphere which is our effective inflaton field $\phi$. Depending on the initial radius of this giant D5 brane, the radius of $S^2$ can either increase or shrink. We will confine ourselves to region $\phi\geq 0$ in the rest of this paper. We analyze each of these regions separately.

\subsection{Region (a)}

In this region, the initial condition for the effective radius of the giant D5 brane is such that it shrinks until it is stabilized at the $\phi=\mu$ vacuum. The total number of e-folds, $N_e$ is
\bea\label{Ne}
M_{P}^{2}\, N_{e}& = & \int_{\phi_{f}}^{\phi_{}{i}} \frac{d \phi \, V}{V'}  \nonumber\\
&=& \frac{1}{8} (y-x) - \frac{\mu^{2}}{32} \ln \left( \frac{\mu^{2} + 4 y}{ \mu^{2} + 4 x  }
 \right)\,,
\eea
 where $\phi_{f}$ and $\phi_i$ are respectively the values of the field at the end and $N_e$ e-folds before the end of inflation. For
convenience we have defined
\be
x\equiv\phi_{f}(\phi_{f} - \mu)\,,\qquad
 y\equiv\phi_{i}(\phi_{i} - \mu)\,.
\ee
The end of inflation  is defined as the point where the first slow-roll parameter,
\be \epsilon= \frac{1}{2} M_{P}^{2}\left( \frac{V'}{V}
\right)^{2}, \ee
becomes equal to one which yields $\phi_f$ such that
\bea
x  = 4 M_{P}^{2} + M_{P} \sqrt{  16 M_{P}^{2} + 2 M_{P}^{2} \mu^{2}     }
\eea
The scalar spectral index at $\phi_{i}$  is
\be
n_{ S}-1=2\eta - 6 \epsilon
\ee
 at $\phi_{i}$   which can be used to eliminate $y$ in favor of $n_S$ and $\mu/M_P$
\be\label{eq-for-mu}
\frac{y}{M_{P}^{2}} =\frac{12+ \sqrt{  144 + 8(1-n_{S}) \frac{\mu^{2}}{M_{P}^{2}}    } }{ 1-n_{S}  }
\ee
Inserting these values for $x$ and $y$ in (\ref{Ne}), we find an
equation for $\mu/M_{P}$ for a given $n_S$. We tried to solve this equation numerically  with
$N_{e}=60$ and $n_S$ between the 2$\sigma$ interval of $n_S$ allowed by Planck,
\be\label{ns-interval}
0.9457\leq n_S\leq 0.9749.
\ee
However not for every value of $n_S$ in this  interval \eqref{ns-interval}, eq.\eqref{eq-for-mu} yields a real solution for $\mu$. The smallest $n_S$, the model can achieve is the one of $\lambda\phi^4/4$ model,
\be
n_S^{\rm min.}=\frac{58}{61}\simeq 0.9508,
\ee
 which could be obtained in the limit of $\mu\rightarrow 0$ of potential \eqref{pot-susy}.
\begin{figure}[t]
\includegraphics[scale=0.6]{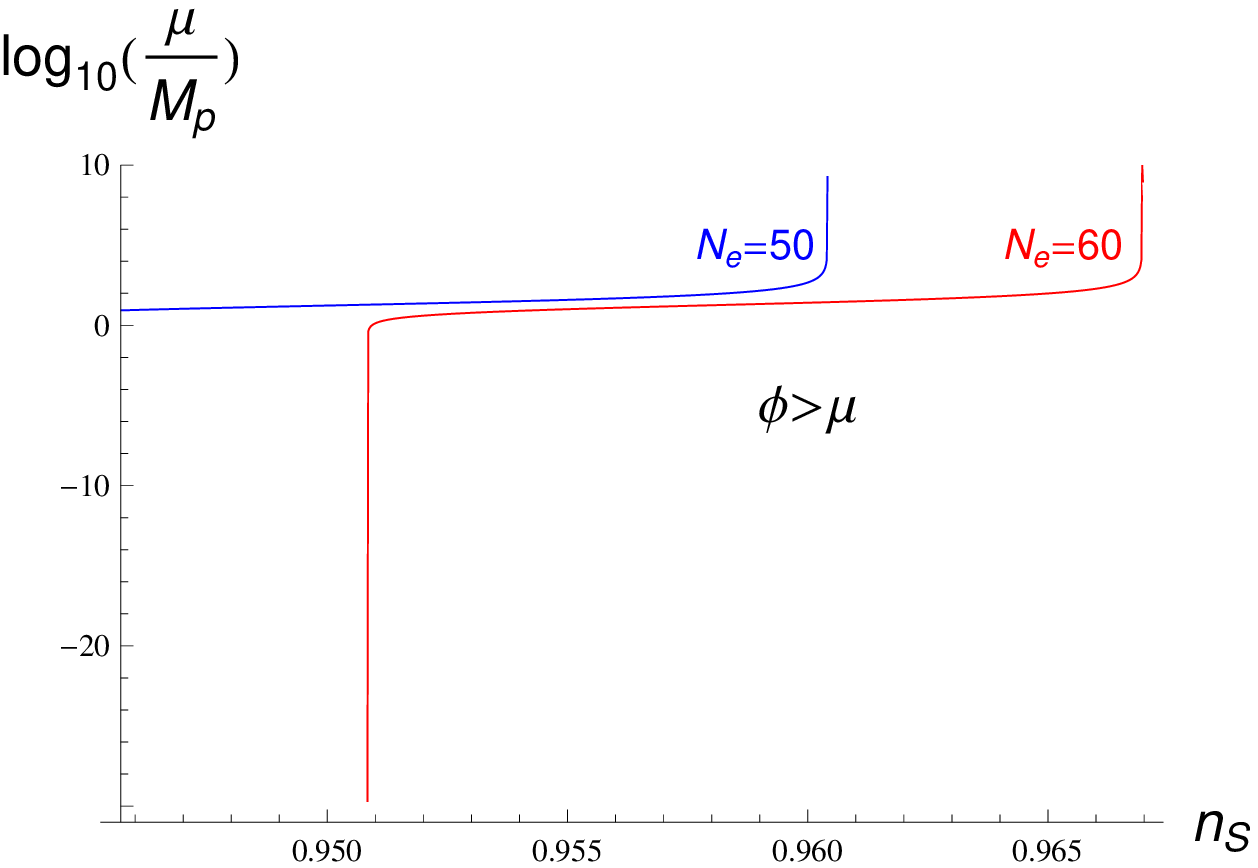}
\includegraphics[scale=0.6]{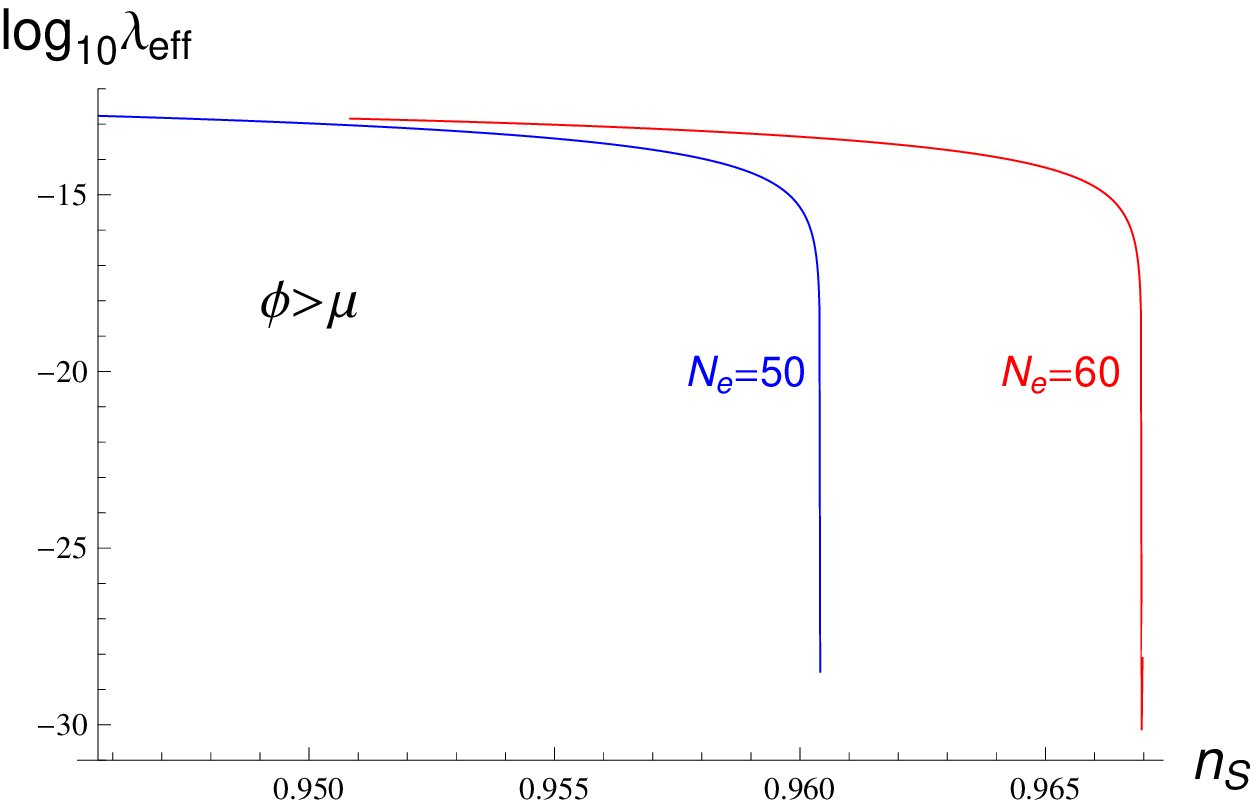}
\caption{Left and right plots respectively demonstrate $\mu$ vs. $n_S$ in the inflationary region where $\phi>\mu$. Blue and red curves are respectively for $N_e=50$ and $N_e=60$. As $\mu\rightarrow 0$, the predictions of the model approaches the $\lambda\phi^4$ model. As $\mu\rightarrow\infty$  $n_S$ reaches its maximal value. For $N_e=50$ and $N_e=50$, the maximum $n_S'$s are respectively $n_S^{50}\lesssim 0.9604$ and $n_S^{60}\lesssim 0.9670$.}
\label{phi-bigger-mu-mu-lambda}
\end{figure}
 It also turns out that for $n_S>n_{S,60}^{\rm max.}$, the equation \eqref{Ne} does not have a real solution for $\mu$. As one can see in the left plot of Fig. \ref{phi-bigger-mu-mu-lambda}, for $n_{S,60}^{\rm max}\simeq 0.9670$, $\mu/M_{p}$ goes to infinity. Thus values of $n_S>n_{S,60}^{\rm max}$ cannot be obtained in this branch of the potential for $N_e=60$. To recap, for $N_e=60$
 \be\label{ns-leg-60}
 0.9508\lesssim  n_S^{60}\lesssim 0.9670.
 \ee
  At the end, the value of $\lambda_{\rm eff}$ could be obtained, matching the amplitude of density perturbations,
 \be
 A_S=\frac{V(\phi_i)}{24\pi^2 \mpl^4 \epsilon(\phi_i)},
 \ee
 with the observed value by Planck, $A_S^{\rm Planck}\simeq 2.1955\times 10^{-9}$. In the right plot of Fig.\ref{phi-bigger-mu-mu-lambda}, one can see how $\lambda_{\rm eff}$ is related to $n_S$. For large values of $\mu/M_P$ when $n_s\rightarrow 0.9670$,  the potential becomes shallower by decreasing the value of $\lambda_{eff}$, {\it i.e.} $\lambda_{eff}\rightarrow 0$. The largest value $\lambda_{eff}$ obtains is for the quartic potential, $\mu=0$,
 \be\label{lambda-max}
 \lambda_{\rm eff}^{60}\lesssim 1.4320\times 10^{-13}.
 \ee

 \begin{figure}[t]
\includegraphics[scale=0.6]{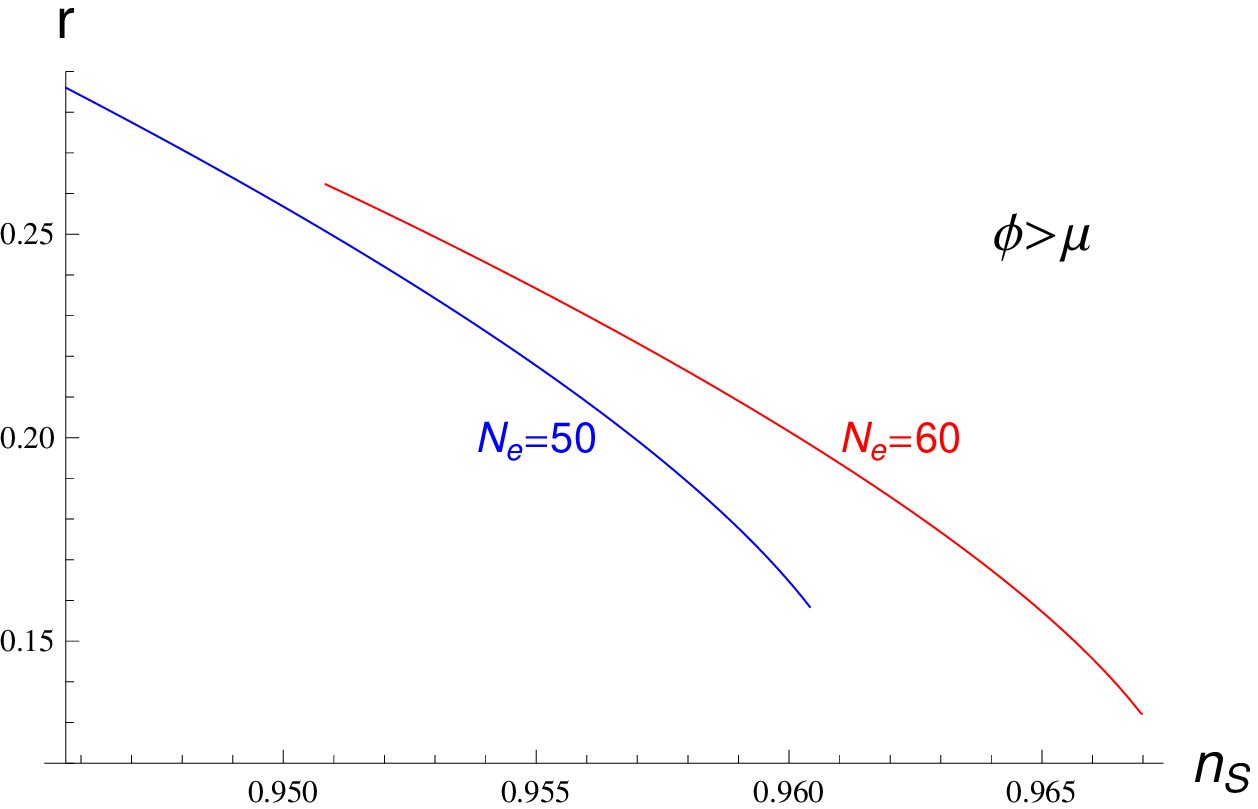}
\includegraphics[scale=0.6]{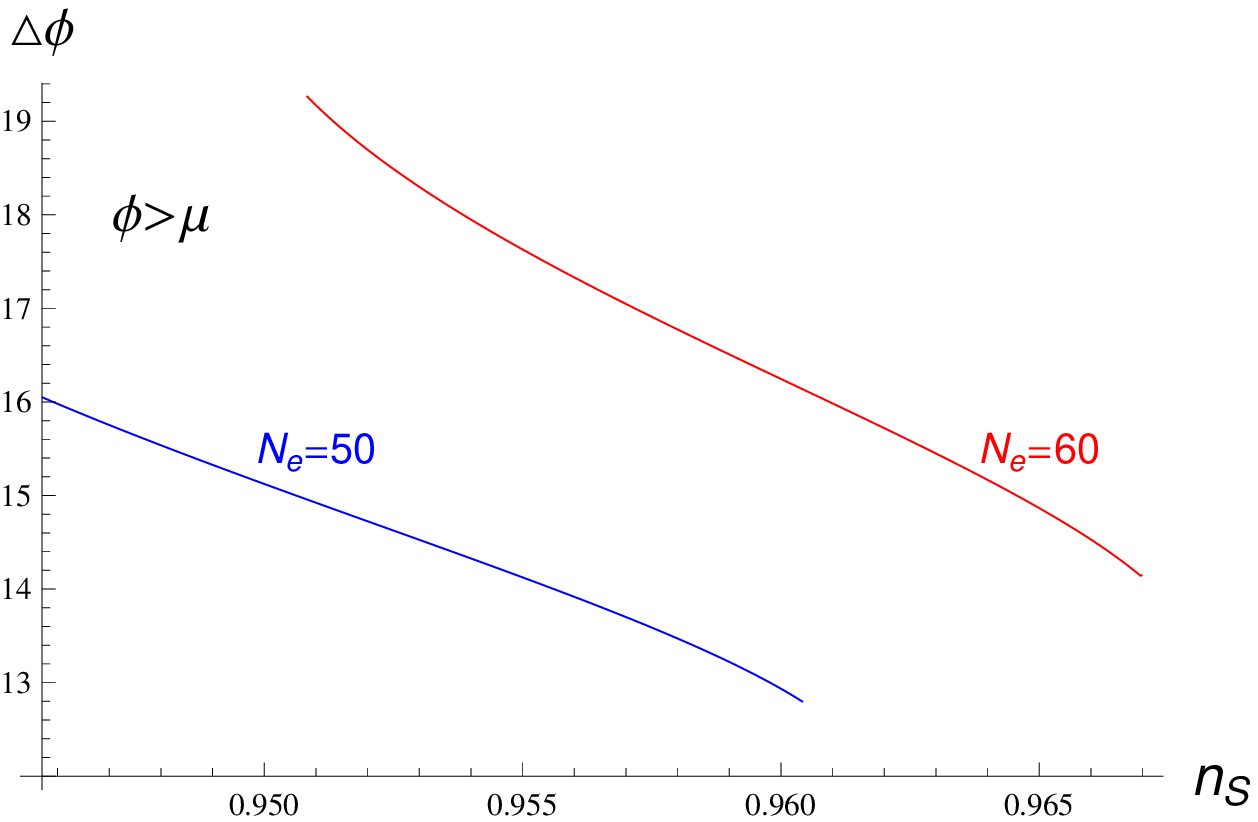}
\caption{Tensor-to-scalar ratio $r$ and field roaming $\Delta\phi$ vs $n_S$ as $n_S$ changes within the allowed ranges for $N_e=50$ and $N_e=60$ when inflation happens in the $\phi>\mu$ region }
\label{r-delta-phi-ns}\end{figure}
In the allowed range of $n_S$ for $N_e=60$, within $2\sigma$ range of the Planck data, the tensor-to-scalar ratio varies in the range
\be
0.1322\lesssim r_{60}\lesssim 0.2623,
\ee
see the left plot in Fig.\ref{r-delta-phi-ns}. It is also interesting to obtain the amount of excursion of the effective inflation for $N_e=60$, $\triangle\phi_{60}$ which turns out to be constrained as
\be
14.15~\mpl\lesssim \triangle\phi_{60}\lesssim 19.26~\mpl.
\ee
Smaller values of $\triangle\phi_{60}$ is obtained for larger values of $n_S$ and $\mu$, see the right plot in Fig.\ref{r-delta-phi-ns}. Even though $\mu\rightarrow \infty$ as $n_S\rightarrow 0.9670$, the displacement of the field decreases and tends to $14.15~\mpl$. In fact the larger the $\mu$, the smaller the amount of the effective inflaton field displacement.

It is expected that in future the $n_S$ is measured with $\sigma(n_S)=0.0029$ \cite{CMBpol} with the CMBPOL experiment. Assuming the central value for $n_S$ remains at its corresponding value from the Planck experiment, $n_S=0.9603$, $r$ still can vary in the range, $r\in \left[0.1983,0.2204\right]$ in the $1\sigma$ range around the central value.

If $N_e=50$, maximum $n_S$ that could be reached is $n_{S,50}^{\rm max.}$, turns out to be $\simeq 0.9604$. For larger values of $n_S$, the equation for $\mu$ does not have any real solution. The lowest value which could be obtained for $N_e=50$ in this potential is again for the $\lambda\phi^4$ theory ($\mu=0$) which is $16/17$. However this value for $n_S$ is already outside the $2\sigma$ region for $n_S$, see \eqref{ns-interval}. Thus we assume that
\be\label{ns-50-efolds}
0.9457\leq n_S\lesssim 0.9604.
\ee
In this range of $n_S$, the quartic coupling and $\mu$ respectively vary such that
\bea\label{mu-lambda-50}
\lambda_{\rm eff}^{50}\leq 9.2945\times 10^{-14}\,,\qquad
\mu_{50} \gtrsim 8.64~\mpl\,.
\eea
The tensor-to-scalar ratio $r_{50}$ and the effective inflaton displacement $\Delta\phi_{50}$  vary as
\be\label{r-delta-phi-50}
0.1584\lesssim r_{50}\leq 0.2508\,,\qquad
12.8~\mpl\lesssim \Delta\phi_{50} \lesssim 15~\mpl
\ee
The smallest value of $\Delta\phi_{50}$ again is obtained when $\mu\rightarrow \infty$. If one allows for the variation of $n_S$ in a one sigma interval then $r_{50}$ varies in $\left[0.1584,0.2341\right]$. If future experiments manage to confine $n_S$ to $\sigma(n_S)=0.0029$, assuming the central value is not changed from the Planck experiment, $r\in\left[0.1584,0.1952\right]$.

In the inflationary region, $\phi>\mu$, the mode with the largest isocurvature amplitude is $j=0$ gauge mode which is massless. Numerical computation of the amplitude of this isocurvature mode suggests that, for $N_e=60$,  $P_{iso}^{A,0}/P_S\lesssim 1.64\times10^{-2}$, which is obtained for the $\mu=0$ case. Our analysis also suggests that with increasing $\mu$ towards infinity, the $P_{iso}^{A,0}/P_S$ goes to $8.27\times10^{-3}$ for $N_e=60$. For $n_S=0.9603$, $P_{iso}^{A,0}/P_S$ is about $1.24\times 10^{-2}$ at the horizon scale. The next isocurvature mode with the largest amplitude is $j=1$ $\beta-$mode which has a constant mass equal to the mass of the inflaton, $m^2$. As $n_S$ varies in the legitimate range, $P_{iso}^{\beta,1}/P_S\lesssim 1.64\times10^{-2}$. The upper bound is again obtained for $\lambda\phi^4$ theory in which $m=\mu=0$. However the ratio $P_{iso}^{\beta,1}/P_S$ goes to $\simeq 3.51\times10^{-5}$. For a given, $n_S$, the $j=1$ $\beta-$mode goes is smaller in comparison with the $j=0$ gauge mode, as it is massive during inflation. For $n_S=0.9603$, $P_{iso}^{\beta,1}/P_S\simeq 4.33\times 10^{-4}$.

\subsection{Region (b) and (c)}

These two regions correspond to the kind of inflationary model which is known as hilltop models \cite{Boubekeur:2005zm}. One can follow the same procedure to obtain the relevant parameters for different values of $n_S$. The predictions of these two regions in the $(n_S,r)$ plane are the same due to the symmetry of the potential around $\phi=\mu/2$. Therefore, we will focus on the case where $\frac{\mu}{2}<\phi_i<\mu$. The only differences with the computations of the previous section are that
 \be
 x=4\mpl^2-\mpl \sqrt{2\mu^2+16{\mpl}^2}
 \ee
and
\be
y=\frac{12-\sqrt{144+8 (1-n_S)\frac{\mu^2}{\mpl^2}}}{1-n_S},
\ee
but the rest of the computations could be applied.

For the central value of $n_S$ from the Planck data, $n_S=0.9603$, for $N_e=60$, the prediction of the model for $r$ is $0.0379$. The symmetry breaking vacuum and the quartic coupling  are respectively, $\mu=33.78\mpl$ and $\lambda_{\rm eff}=7.0306\times 10^{-14}$.  Again for $N_e=60$, one finds out that within the $2\sigma$ range of $n_S$ allowed by the Planck experiment, the equation for $\mu$ does not have any real solution for $n_{S,60}^{\rm max.}\gtrsim 0.9681$.
Thus we assume
\be\label{ns-60-phi-less-mu}
0.9457<n_S^{60}<0.9681
\ee
 \begin{figure}[t]
\includegraphics[scale=0.6]{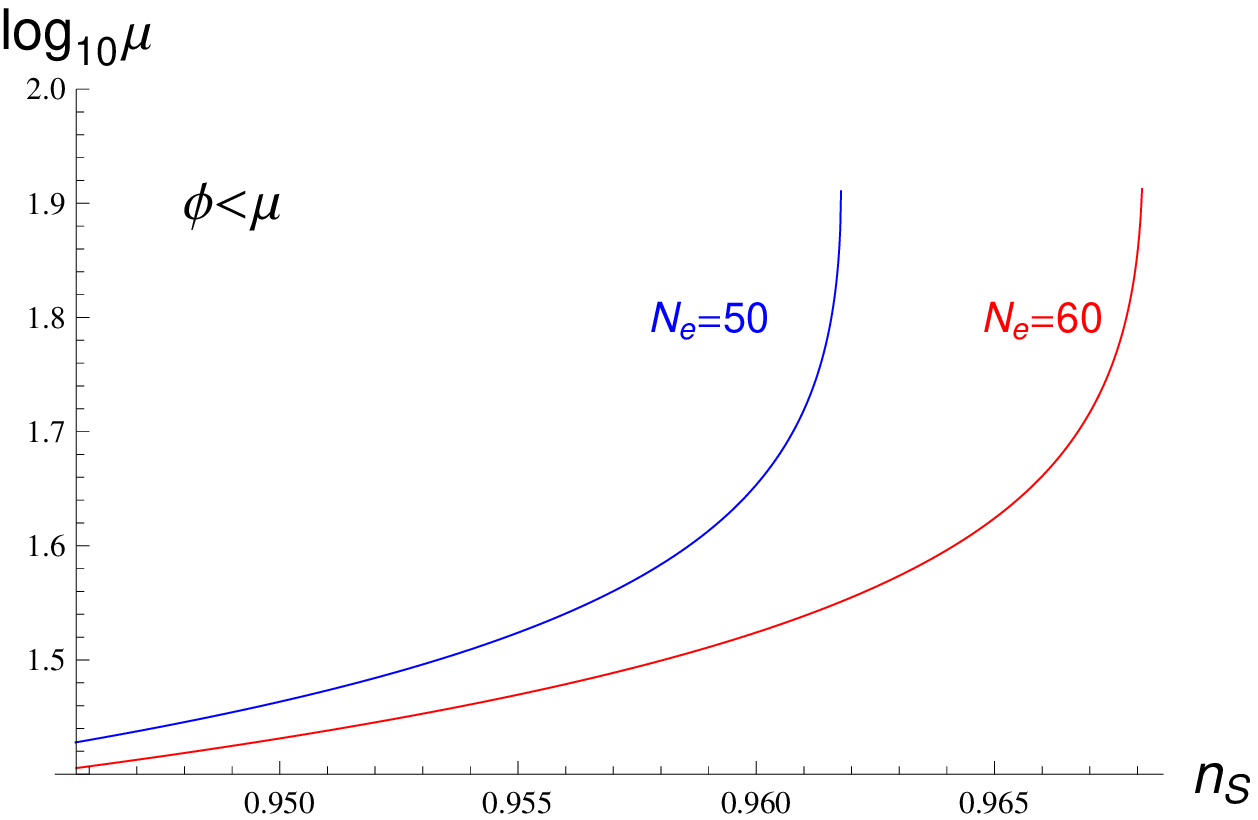}
\includegraphics[scale=0.6]{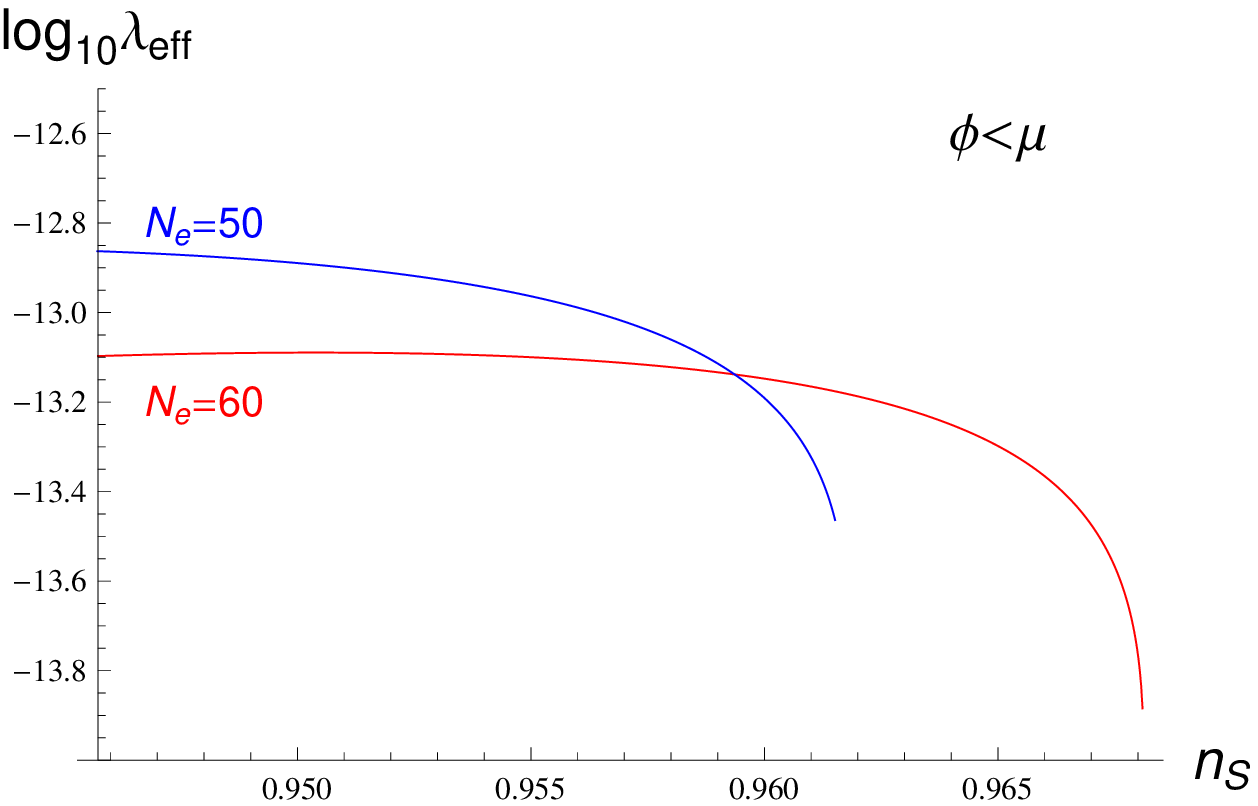}
\caption{$\mu$ and $\lambda_{\rm eff}$ vs. $n_S$ in the region $\phi<\mu$}
\label{phi-less-than-mu-lambda-mu}
\end{figure}
 As can be seen from the Fig.\ref{phi-less-than-mu-lambda-mu}, $\lambda_{\rm eff}$  and $\mu_{60}$ in this range of $n_S$ varies in the small interval
\be
1.3053 \times 10^{-14}\lesssim \lambda_{\rm eff}^{60}\lesssim 7.9948\times 10^{-14}\,,\qquad
25.43~\mpl \lesssim\mu_{60}\lesssim 81.76~\mpl.
\ee

\begin{figure}[t]
\includegraphics[scale=0.6]{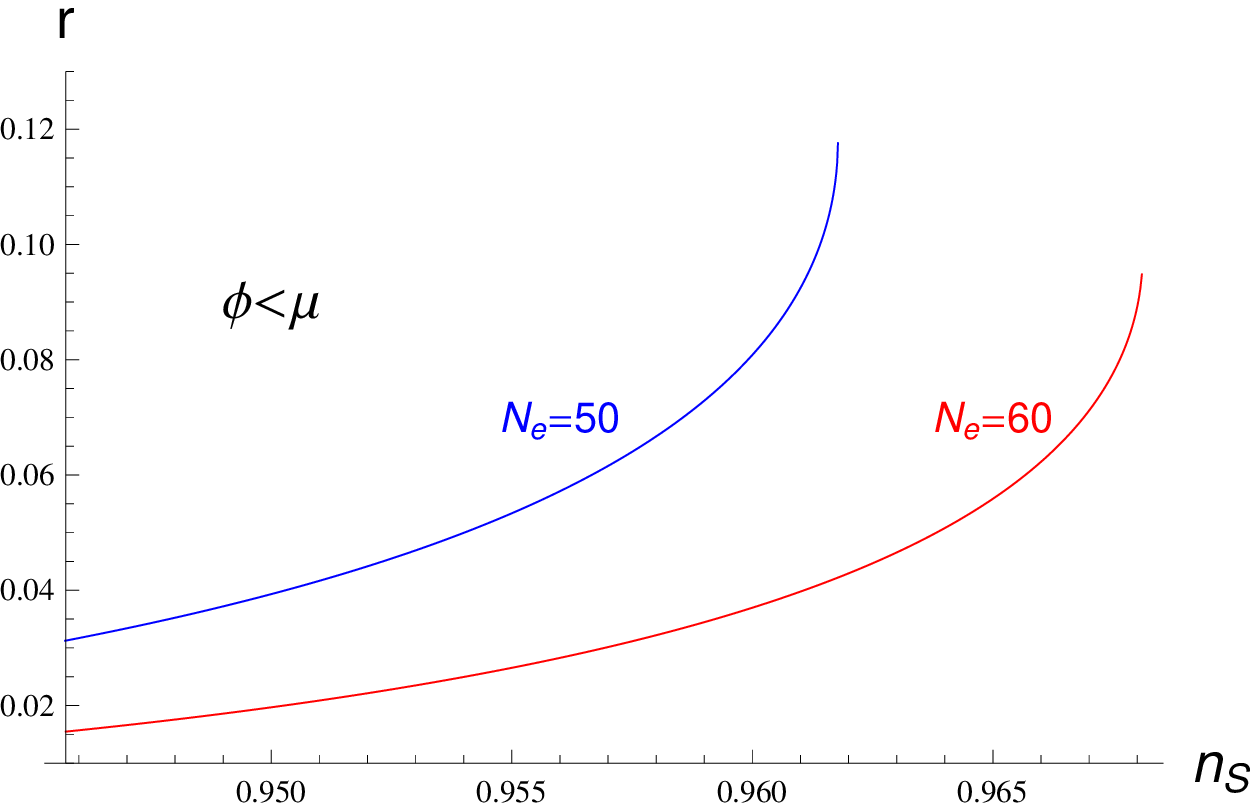}
\includegraphics[scale=0.6]{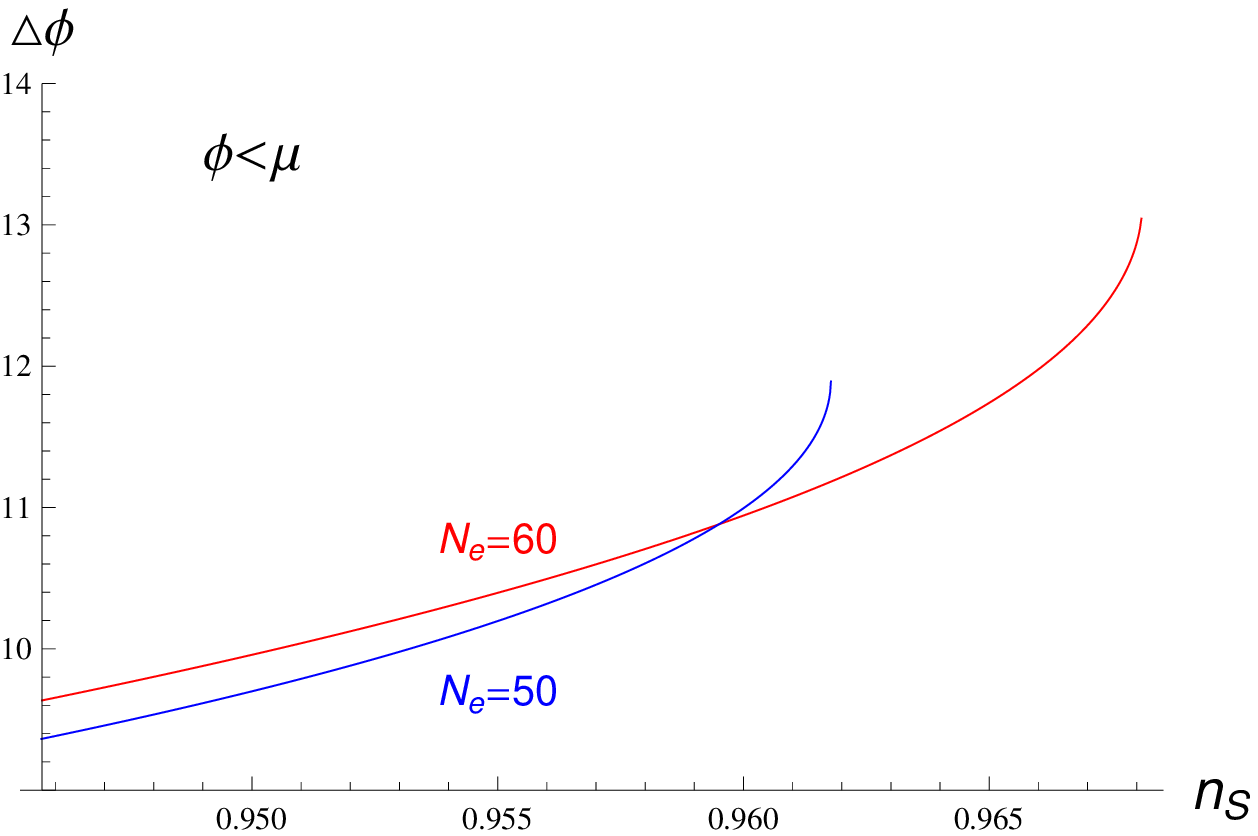}
\caption{$r$ and $\Delta\phi$ vs. $n_S$ in the region $\phi<\mu$.}
\label{phi-less-than-mu-r-dphi}
\end{figure}

The tensor-to-scalar ratio $r_{60}$ and the field displacement $\Delta\phi_{60}$  vary in the following ranges,
\bea\label{r-60-phi-less-mu}
0.0155\lesssim r_{60}\lesssim 0.0948\,,\qquad
25.43~\mpl\lesssim \Delta\phi_{60}\lesssim 81.76~\mpl,
\eea
please see Fig. \ref{phi-less-than-mu-r-dphi}. If $\sigma(n_S)$ could be lowered to $0.0029$, as \cite{CMBpol} suggests, assuming the central value for $n_S$ remains as in the Planck experiment, $r$ varies in $\left[0.0310,0.0475\right]$.

If $N_e=50$, the maximum $n_S$ that could be obtained in this region is $n_{S,50}^{\rm max.}\simeq 0.9618$. For larger values of $n_S$, the equation for $\mu$ does not have any real solution. Varying $n_S^{50}$ between $0.9457$, the lower bound of the Planck experiment $2\sigma$ region, and $n_{S,50}^{\rm max.}$, \emph{cf}. Fig. \ref{phi-less-than-mu-lambda-mu},
\bea
1.3717 \times 10^{-14}\lesssim \lambda_{\rm eff}^{50}\lesssim 1.8678\times 10^{-14}\,,\qquad
26.78~\mpl \lesssim\mu_{50}\lesssim 81.4~\mpl.
\eea
The tensor-to-scalar ratio $r_{50}$ and the field displacement $\Delta\phi_{50}$ vary in the following ranges, (\emph{cf}. Fig. \ref{phi-less-than-mu-r-dphi})
\bea\label{r-50-phi-less-mu}
0.0312\lesssim r_{50}\lesssim 0.1176\,,\qquad
9.36~\mpl\lesssim \Delta\phi_{50}\lesssim 11.89~\mpl,
\eea
If $\sigma(n_S)$ is lowered to $0.0029$ still the predictions of the model for $r$ varies between $0.0636$ and $0.1176$.

Even though the predictions of regions (b) and (c) are the same in the $(n_S,r)$ plane, their forecast for the amplitude of isocurvature perturbations at the end of inflation are in general different, as the masses of the spectator modes are in general $\phi-$dependent. The lightest mode in region (b) is again the $j=0$ gauge mode. For $60$ e-folds, increasing $n_S$ from $0.9457$ to its maximum $0.9681$, $P_{iso}^{A,0}/P_S$ increases approximately from $9.83\times 10^{-4}$ to $6.04\times 10^{-3}$. The next isocurvature mode with the largest amplitude in the tower is $j=1$ $\beta$ mode whose amplitude decreases from $7.67\times10^{-6}$ to $5.96\times 10^{-9}$ as $n_S$ increases in the allowed range.

In region (a), the mode with the largest amplitude for isocurvature perturbations is $j=0$ $\alpha-$mode. Its amplitude $P_{iso}^{\alpha,1}/P_S$ varies between $2.91\times 10^{-2}$ and $2.84\times 10^{-4}$. The next ones are $j=0$ gauge and $j=1$ $\beta-$modes which have the same amplitude as region (b) due to the symmetry of their mass with respect to the symmetry $\phi\rightarrow \mu-\phi$. In region (a), the model is equipped with an embedded preheating mechanism that uses the isocurvature fields as preheat modes. The couplings of the preheat modes to the inflaton are related to the inflaton's self-couplings hence known. In \cite{M-flation-Preheating-GWs}, we numerically computed the amplitude of gravitational waves which is generated during preheating. We found that the spectrum peaks in the GHz band with  $\Omega_{\rm GW} h_0^2 \propto 10^{-16}$.

\section{Concluding remarks}\label{discussion-section}\label{discussion-sec}

M-flation is a many-field inflationary model that could be realized within string theory from the dynamics of a stack of D-branes exposed to the six dimensional flux configuration. Due to the Myers effect, two of the dimension perpendicular to the D3 branes puffs up to a fuzzy sphere whose radius plays the role of inflaton. Requiring the 10 dimensional background to be a solution to the supergravity equations, forces the shape of the potential to be a symmetry-breaking double-well potential. In the introduction we summarized some specific theoretical features of M-flation and its theoretically appealing features. Here we point out two such other features:

\paragraph{1) Sub-Planckain field displacements.} We note that the individual physical fields displacements during our M-flation is $10^{-6} M_p\ll M_p$. Such displacements can even be  produced by sub-Planckian thermal pre-inflation fluctuations which ``collectively'' give rise to an ``effective" super-Planckian displacement needed for the ``effective'' inflaton. This is different from usual chaotic models in which the field should take super-Planckian v.e.v.'s which could only be achieved anthropically  or if the thermal bath before inflation has (super-)Planckian energy densities.

\paragraph{2) Stability of inflationary trajectories.} Given large number of spectator modes one may worry about their backreaction destroying the slow-roll trajectories. One can distinguish two such effect, one is coming from the isocurvature modes which have crossed the horizon and became classical and hence have a non-zero v.e.v which may backreact on inflationary trajectory; or consider the effects of sub-horizon quantum modes and the deformation of the potential due to quantum loop effects. The latter was discussed in detail in \cite{M-flation-1} and discussed that basically due to the shape of potential (induced from supersymmetric 10d background) such quantum corrections are small. The former, was also discussed in \cite{gauged-M-flation,M-flation-1}, and here we give a short review of the discussion:  The power spectrum of the spectators (which is basically a measure of the energy carried by these modes and hence a measure of their backreaction) drops down exponentially with their mass as $\exp(-m/H)$. On the other hand, the mass of states grows with quantum number $j$, which also controls their degeneracy $2j+1$. Therefore, the growth of degeneracy is basically overshadowed by the exponential suppression of the modes due to their mass and in the end the effects of these isocurvature modes remain small. (Recall that as we already discussed the power spectrum of lightest isocurvature modes is of order $10^{-2}$ less than the power spectrum of curvature perturbations.) Altogether the total backreaction of the isocurvature modes remains small and doe not undermine M-flation slow-roll trajectories.


We also discussed phenomenological appealing features of gauged M-flation with double-well potential.
In the region $\phi>\mu$  predictions of the model is at the central point of the BICEP2 data, $r=0.2$ \cite{BICEP-data}, assuming that $n_S$ is at the central value Planck yields for the scalar spectral index, $n_S=0.9603$. Due to the hierarchical structure of the mass spectrum of the spectator species, the model is immune to super-cutoff excursion problem, which undermines N-flation \cite{gauged-M-flation}. Also in the $\phi>\mu$ region, the Hessian mass matrix of all spectator modes is a positive-definite,  suggesting that the SU(2) sector in this direction is an attractor. SU(2) sector is also an attractor in the hilltop inflationary region $\mu/2<\phi<\mu$. Since the inflationary potential can in principle approach the Planckian energy densities in the $\phi>\mu$ region, the energy density in the inflationary Planck length patch at that epoch can have a vacuum energy comparable to the energy density of the anisotropies and thus can dilute them. Thus it is more likely to lead to inflation, which justifies the large value of $r$ observed by the BICEP2 region.

In the hilltop region $0<\phi<\mu/2$ the first $\simeq 93$ modes become tachyonic, at least for a while, during the $60$ e-folds of inflation \cite{gauged-M-flation}. However the mass squared of these modes, despite of being tachyonic, remain of order $m^2\simeq -0.01 H^2$, where $H$ is the inflationary Hubble parameter during inflation. This suggests that even though the inflaton may roll off the SU(2) direction along those unstable ones, it may lead to a hilltop-like inflation along those orthogonal direction. We postpone the thorough analysis of this point to future works.

In this note, we analyzed the predictions of M-flation in different inflationary regions assuming that $n_S$ varies in the $2\sigma$ region allowed by the Planck data \cite{Planck-data}. Even though the $2\sigma$ region is small, we notice that in different inflationary regions some ranges of $n_S$ might not be approached by the model. For $N_e=60$, values of $0.9508\lesssim n_S\lesssim 0.9670$ could be achieved. The minimum value for $n_S$ is obtained for $\mu=0$, \ie the quartic chaotic model. Values close to $0.9670$ is obtained for $\mu\rightarrow \infty$. The field displacement still remains of order $\mathcal{O}(10)\mpl$ despite $\mu$ becoming large. What happens is that $\lambda_{\rm eff}$ tends towards zero as $\mu$ increases. Nonetheless, if one demands to avoid the ``hyper-Planckian'' v.e.v's for $\mu$, say $\mu\lesssim 100 \mpl$, maximum $n_S$ will reduce to $0.9650$. Another potentially observable signature of the model in the region $\phi>\mu$, is a small isocurvature perturbation mode with amplitude of $\simeq 10^{-2}$ which is obtained from $j=0$ gauge mode. Of course whether this mode would be observable, depends on the details of the reheating.

Although BICEP2 claims to observe a B-mode signature corresponding to $r_{0.05}=0.2_{-0.05}^{+0.07}$, the Planck data so far has only suggested an upper bound on the amplitude of tensor modes $r_{0.002}<0.11$. There has been various proposals to patch up the BICEP2 results with the Planck upper bound on the B-modes \cite{BICEP-Planck-Reconcile}. In \cite{non-BD-2}, based on our previous analysis in \cite{non-BD-1} for excited states, we proposed a scenario to reconcile the large scale models of inflation with the Planck data and BICEP data. For that, one has to assume that the scale of new physics is about $M=\mathrm{few}\times 10 H$. To reconcile the Planck and BICEP2 data, we proposed two approaches to solve the tension between the Planck and BICEP2 data:
i) creating a running scalar spectral index with running of order $d n_S/d\ln k\simeq -0.02$, or ii) creating a blue tensor spectrum for the tensor modes.
 Both these two approaches could be easily accustomed to M-flation assuming non-Bunch-Davis, excited initial state for cosmic perturbations.

\section*{Acknowledgements}

The authors are thankful to D. Lyth and Robert Mann for useful comments. A.A. is supported by the Lancaster-Manchester-Sheffield Consortium for Fundamental Physics under STFC grant ST/J000418/1. M.M.Sh-J would like to acknowledge the International Visiting Scholar Program of Kyung Hee University Seoul.

\end{document}